\documentclass[11pt]{article}
\usepackage{moriond,epsfig}
\usepackage{graphicx,changebar}
\usepackage{amsmath}
\def\leqsim{\mathbin{\;\raise1pt\hbox{$<$}\kern-8pt\lower3pt\hbox{$\sim$}\;}}
\def\geqsim{\mathbin{\;\raise1pt\hbox{$>$}\kern-8pt\lower3pt\hbox{$\sim$}\;}}

\def\MXN#1{\mbox{$ M_{\tilde{\chi}^0_#1}                                $}}

\def\XP#1{\mbox{$ \tilde{\chi}^+_#1                                     $}}
\def\XM#1{\mbox{$ \tilde{\chi}^-_#1                                     $}}
\def\XPM#1{\mbox{$ \tilde{\chi}^{\pm}_#1                                $}}
\def\XN#1{\mbox{$ \tilde{\chi}^0_#1                                     $}}

\def\p#1{\mbox{$ \mbox{\bf p}_1                                         $}}

\newcommand{\tanb}    {\mbox{$ \tan \beta                                  $}}

\newcommand{\snu}     {\mbox{$ \tilde\nu                                   $}}
\newcommand{\msnu}    {\mbox{$ m_{\tilde\nu}                               $}}

\newcommand{\sfe}     {\mbox{$ \tilde{\mathrm f}                           $}}

\newcommand{\stq}     {\mbox{$ \tilde {\mathrm t}                          $}}
\newcommand{\stqone}  {\mbox{$ \tilde {\mathrm t}_1                        $}}

\newcommand{\mstqone}    {\mbox{$ M_{\tilde {\mathrm t}_1}                 $}}

\newcommand{\hn}      {\mbox{$ {\, \mathrm h}^0                               $}}

\newcommand{\sqs}     {\mbox{$ \sqrt{s}                                    $}}

\newcommand{\ee}      {\mbox{$ {\, \mathrm e}^+ {\mathrm e}^-                 $}}

\newcommand{\eeto}    {\mbox{$ {\, \mathrm e}^+ {\mathrm e}^- \to             $}}

\newcommand{\GeV}     {\mbox{$ {\mathrm{GeV}}                              $}}

\newcommand{\GeVcc}   {\mbox{$ {\mathrm{GeV}}/c^2                          $}}

\newcommand{\pbi}     {\mbox{$ {\mathrm{pb}}^{-1}                          $}}

\newcommand{\ffbarp}  {\mbox{$ {\, \mathrm f} \, \bar{\mathrm f}'          $}}

\newcommand{\charm}   {\mbox{$\mathrm c                                    $}}
\newcommand{\bottom}  {\mbox{$\mathrm b                                    $}}
\newcommand{\topq}    {\mbox{$\mathrm t                                    $}}

\def\PR#1#2#3{{\rm Phys.~Rep.} {\bf#1} (#2) #3}

\def    \DM          {\mbox{$\Delta$M}}
\def    \missEt      {\ifmmode{/\mkern-11mu E_t}\else{${/\mkern-11mu E_t}$}\fi}
\def    \missE       {\ifmmode{/\mkern-11mu E}\else{${/\mkern-11mu E}$}\fi}
\def    \missp       {\ifmmode{/\mkern-11mu p}\else{${/\mkern-11mu p}$}\fi}
\def    \misspt      {\ifmmode{/\mkern-11mu p_t}\else{${/\mkern-11mu p_t}$}\fi}

\def\be{\begin{equation}}
\def\ee{\end{equation}}
\def\bea{\begin{eqnarray}}
\def\eea{\end{eqnarray}}

\begin{document}
\title{MSSM SUGRA searches at LEP}

\author{T. Alderweireld}

\address{Facult\'e des Sciences, Universit\'e de l' Etat Mons, Mons, Belgium \\ email: thomas.alderweireld@cern.ch}

\maketitle\abstracts{During the LEP2 operation, the 4 LEP experiments have collected data at centre--of--mass energies up to 209 \GeV . Those data have been analysed 
in the search of charginos, neutralinos and sfermions in the framework of the Minimal Supersymmetric Standard Model (MSSM) with R--parity conservation and assuming the 
lightest neutralino to be the Lightest Supersymmetric Particle (SUGRA--model). No evidence for a signal was found in any of the channels. The results of each 
search were used to derive upper limits on production cross-sections and masses. In addition, the combined result of all searches excludes regions 
in the parameter space of the constrained MSSM, leading to limits on the mass of the LSP. All limits are given at 95\% confidence level, all results are preliminary.}

\section{Introduction}\label{sec:intro}
Supersymmetry (SUSY)~\cite{susy} is at present one of the most attractive possible extensions of the Standard Model (SM) and its signatures 
could in principle  be observed at LEP through a large  variety of different channels. Searches for pair-produced charginos, neutralinos, sleptons and squarks
are briefly reviewed. The searches were performed and interpreted in the most model-independent  possible way in terms of production cross-sections and masses. 
The results are interpreted in the framework of SUSY models, with different search channels complementing each other in constraining the parameter space.
During the  LEP2 operation (1995-2000), the 4 LEP experiments, ALEPH, DELPHI, L3 and OPAL have each collected  about 700 \pbi\ of integrated luminosity at 
centre--of--mass energies up to 209 \GeV . Those data have been used for the present searches.

\section{SUSY framework}\label{sec:frame}
The searches presented here were performed in the framework of the Minimal Supersymmetric extension of the Standard Model
(MSSM)~\cite{susy}. $R$-parity conservation is assumed, implying that the Lightest Supersymmetric Particle (LSP) is stable and SUSY particles (``sparticles'', defined as having $R=-1$) are pair-produced. 
In addition, they decay directly or indirectly into the LSP which is weakly interacting and escapes detection, giving a signature of missing energy and momentum.
The lightest neutralino ($\tilde{\chi}^0_1$) is assumed to be the LSP (SUGRA--model). To make the model more predictive, the unification of some parameters at a high mass scale typical 
of Grand Unified Theories (GUT) is assumed, leading to only 8 new parameters with respect to the SM ones~\cite{susy,results} namely   $\mu$, $m_{\frac{1}{2}}$, $m_{0}$, \tanb\ , $m_A$ and  $A_{b,t,\tau}$. 

\section{Results}\label{sec:res}
The preliminary results given here consist of a selection of the searches fully reviewed in~\cite{results}. In any of the searches no excess of candidates was observed 
with respect to the standard model predictions and limits on masses and cross--sections were set at 95 \% confidence level.

\subsection{Search for squarks and sleptons}\label{subsec:sqa}
In large regions of the SUSY parameter space the dominant decay of the sfermions is
to the corresponding fermion and the lightest neutralino, 
$\sfe \rightarrow {\mathrm f} \XN{1} $. 
In the case of the \stq, the decay 
$\stq \rightarrow \topq \tilde{\chi}^0_1 $ 
is not kinematically allowed at LEP, and the dominant 2-body decay channel is expected
to be $\stq \rightarrow \charm \XN{1} $  
($\stq \rightarrow \bottom \XPM{1} $ being disfavoured by existing
limits on the chargino mass). If $\msnu\! <\! \mstqone $, the three-body
decay $\stqone\! \to$ b$\ell\snu$ may compete with the c\XN{1} decay.

Therefore, final state topologies with a pair of leptons or jets and missing energy are the 
relevant ones in the search for sleptons and squarks, respectively. The 
total energy of the detectable final state particles 
(and, thus, the sensitivity of the search)
depends primarily of the mass difference between the sfermion and the LSP (generically called \DM ).

In the third slepton and squark famillies mass eigenstates, ($\tilde{f}_{1}$,$\tilde{f}_{2}$) may be a mixing between interaction eigenstates 
($\tilde{f}_{L}$,$\tilde{f}_{R}$), this mixing is usually labelled by a $\theta_{mix}$\ mixing angle which depends on the theoretical parameter values of the SUSY model.
For some peculiar values of this angle a decoupling to the Z boson may occur and the total production cross-section decreases and so is the corresponding mass limit.
Table~\ref{tab:sqar} shows the combined results of the 4 LEP collaborations in term of the obtained squark mass limits at 95\% C.L. 
requiring \DM\ to be greater than 15 \GeVcc and the ALEPH $\tilde{t}_{1}$ mass limit valid for all \DM\ values. Only limits on the third squark familly are shown
since $\tilde{t}_{1}$ and $\tilde{b}_{1}$ are expected to be lighter than the squarks from the two first famillies due to the non-diagonal mixing terms in the Lagrangian.
For the sleptons, table~\ref{tab:slept} shows the obtained mass limit for $\tilde{e}_{R}$, $\tilde{\mu}_{R}$ and $\tilde{\tau}_{R}$ combining all LEP2 data. 

\begin{table}[ht]
\begin{center}
\begin{tabular}{|l|c|c|}
\hline
{Limits (ADLO)}     & \multicolumn{2}{c|}{{$\Delta M >$~15~GeV/$c^2$}} \\
\hline
 & $\theta_{mix} = 0^{\circ} $  &  $\theta_{mix} = 56^{\circ}/68^{\circ}$ \\
\hline
\it{ {stop} mass limit} $\tilde{t}_{1} \rightarrow c   \tilde{\chi}$ &  $>$ 96~GeV/$c^2$ &  $>$ 93~GeV/$c^2$ \\
\hline
\it{ {stop} mass limit} $\tilde{t}_{1} \rightarrow b l \tilde{\nu} $  &  $>$ 95~GeV/$c^2$ & $>$ 91~GeV/$c^2$   \\
\hline
\it{ {sbottom} mass limit} $\tilde{b}_{1} \rightarrow b   \tilde{\chi}$ &  $>$ 99~GeV/$c^2$ &  $>$ 93~GeV/$c^2$  \\
\hline
\multicolumn{3}{|c|}{{ALEPH limit for all $\Delta M$, using long-lived hadron search }} \\
\hline
\it{ {stop} mass limit} $\tilde{t}_{1} \rightarrow c   \tilde{\chi}$ & \multicolumn{2}{c|}{ $>$ 65~GeV/$c^2$}  \\
\hline
\end{tabular}
\end{center}
\caption[.]{
\label{tab:sqar}
Mass limits on squark masses obtained with a combination of the 4 LEP experiments requiring $\DM >$~15~\GeVcc~.
The last line gives the mass limit obtained by ALEPH where the \DM\ constraint was~released.}
\end{table}

\begin{table}[ht]
\begin{center}
\begin{tabular}{|l|c|}
\hline
{Limits (ADLO)}     &   { $\Delta M >$~15~GeV/$c^2$} \\
\hline
\it{ { selectron} mass limit} &  $>$ 99~GeV/$c^2$  \\
\hline
\it{ { smuon}     mass limit} &  $>$ 94~GeV/$c^2$  \\
\hline
\it{ { stau}      mass limit} &  $>$ 80~GeV/$c^2$  \\
\hline
\end{tabular}
\end{center}
\caption[.]{
\label{tab:slept}
Obtained mass limit for $\tilde{e}_{R}$, $\tilde{\mu}_{R}$ and $\tilde{\tau}_{R}$ combining all LEP2 data and requiring $\DM > $ 15 \GeVcc . }
\end{table}

\subsection{Search for charginos and neutralinos}\label{subsec:char}

In the case of chargino pair production, the final state is 
four jets if both charginos decay hadronically, two jets and one lepton if
one chargino decays into $\ell \nu \XN{1}$, and leptons only if both charginos
decay into leptons. The branching-ratio of $\XPM{1} \to \XN{2} \ffbarp$ can 
be sizeable, in particular in the regions of the parameter space where 
 $\XN{2} \rightarrow \XN{1} \gamma$ is important. In this case, the above
topologies are accompanied by photons. 

If the mass difference \DM\ between the chargino and the LSP is very small 
the visible energy released in the decay is very small, making the 
signal hard to detect. The simultaneous production of a photon by initial state radiation was used to
explore such regions, as this allows a very efficient background rejection at 
the expense of a low signal cross-section.
Even lower mass differences imply a long lifetime of the chargino, which can then be
identified as a heavy stable charged particle or one with a displaced decay
vertex.

In the case of the detectable \XN{1}\XN{k} neutralino production channels 
($i.e.$ excluding \XN{1}\XN{1}), the most important signatures are 
expected to be acoplanar pairs of jets or leptons with high missing energy and momentum. 

\begin{figure}[h]
\begin{minipage}{.49\linewidth}
\epsfysize=7.0cm\epsffile{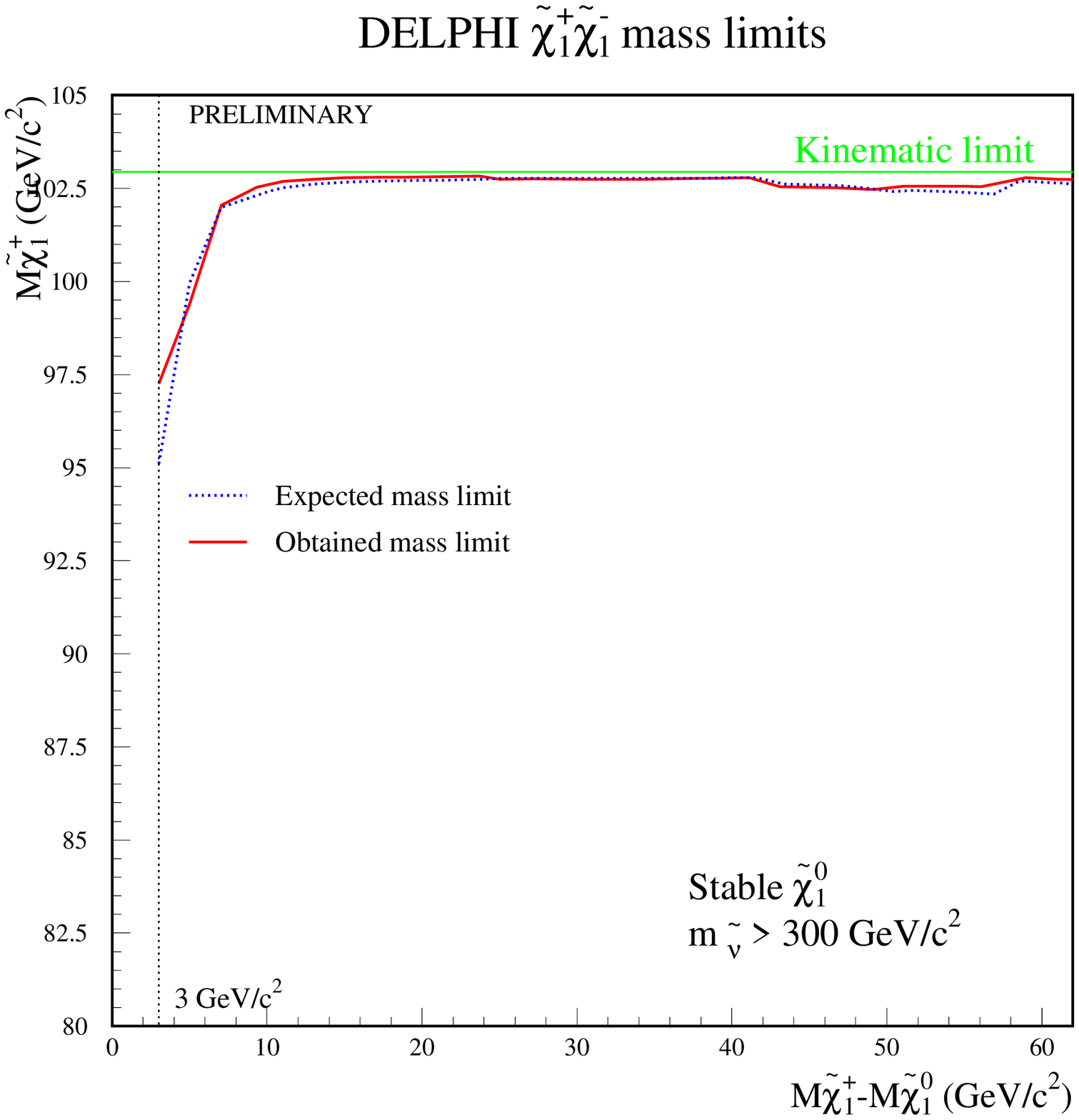}
\caption{The chargino mass limit as function of the \DM\ value
under the assumption of a heavy sneutrino.
The limit applies to the
case of a stable \XN{1}.
The straight horizontal line shows the kinematic limit.
\label{fig:chamass}}
\end{minipage}\hfill
\begin{minipage}{.49\linewidth}
\epsfysize=7.0cm\epsffile{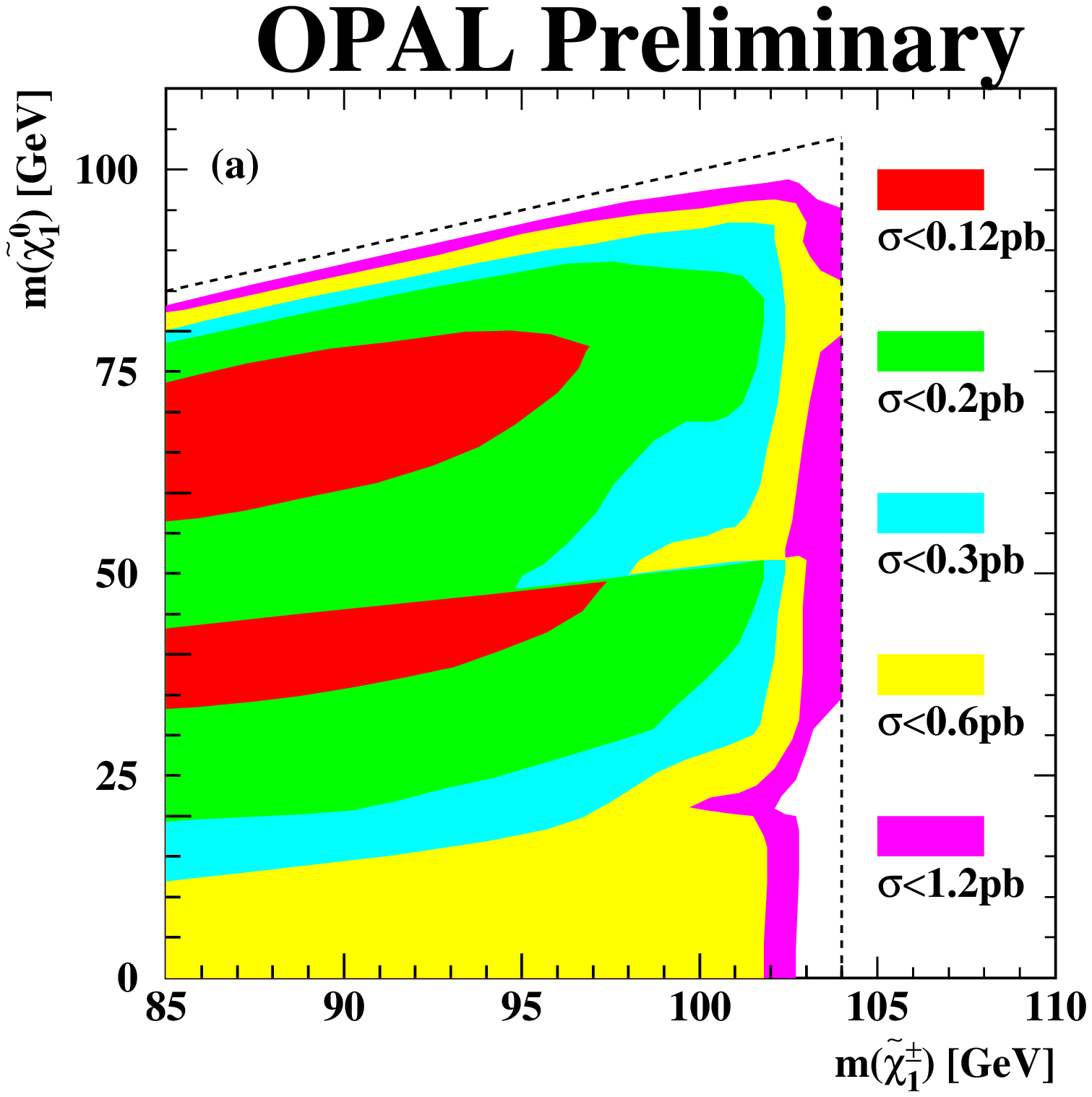}
\caption{The contour of the 95\% C.L. upper limits for the \eeto\XP{1}\XM{1}\ 
production cross--sections at \sqs\ = 208 \GeV\ are shown assuming 
Br($\XN{2}\rightarrow Z^* \XN{1}$) = 100 \%  The kinematic boundaries are shown 
by dashed lines.
\label{fig:chacross}}
\end{minipage}
\end{figure}

\begin{figure}[h]
\begin{minipage}{.49\linewidth}
\epsfysize=7.0cm\epsffile{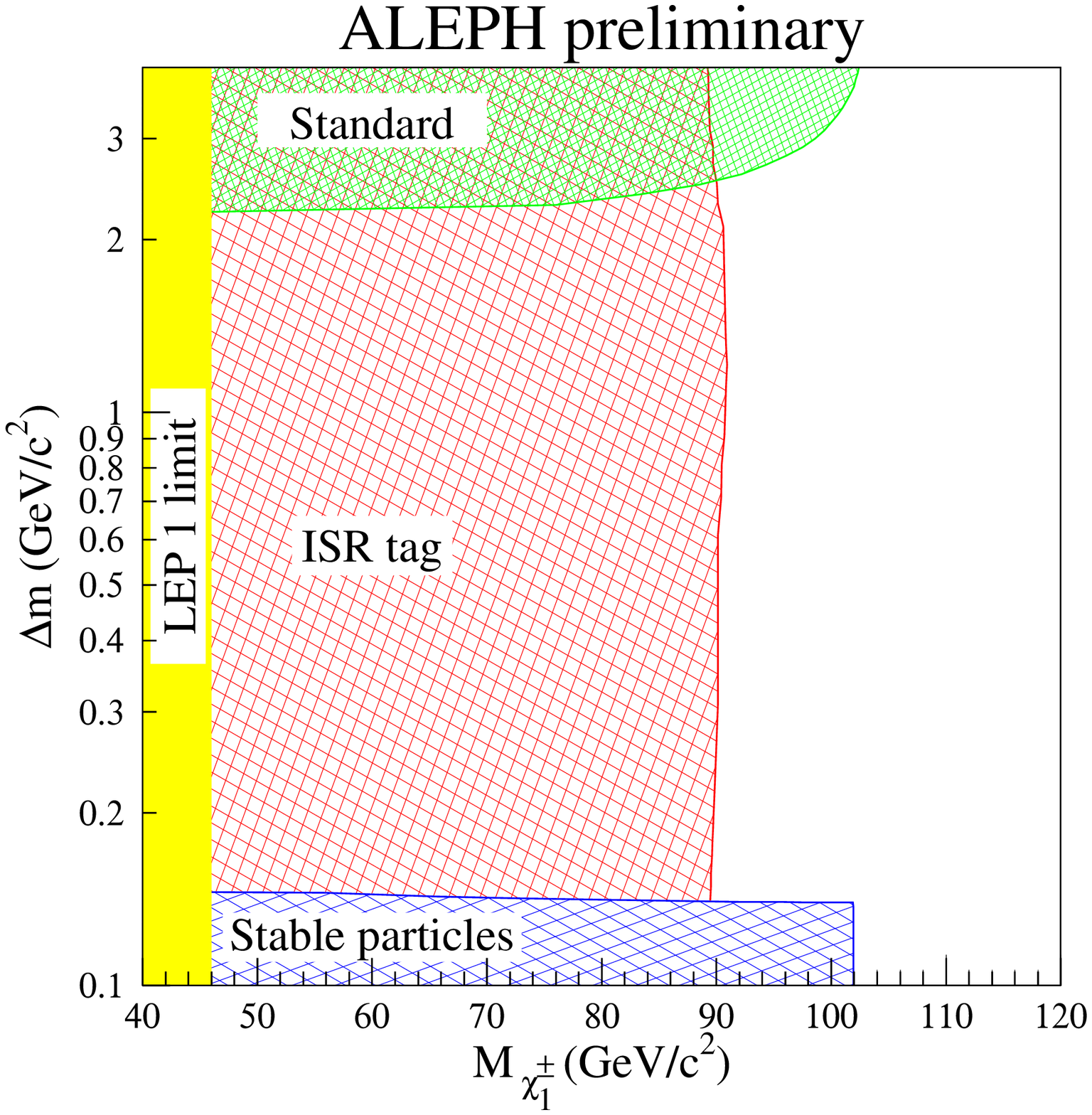}
\caption{95\% confidence level excluded region for $m_0>$~500~\GeVcc\ in the higgsino region.
Chargino masses below 89 \GeVcc\ are excluded at 95\% C.L. 
\label{fig:chadeg}}
\end{minipage}\hfill
\begin{minipage}{.49\linewidth}
\epsfysize=6.5cm\epsffile{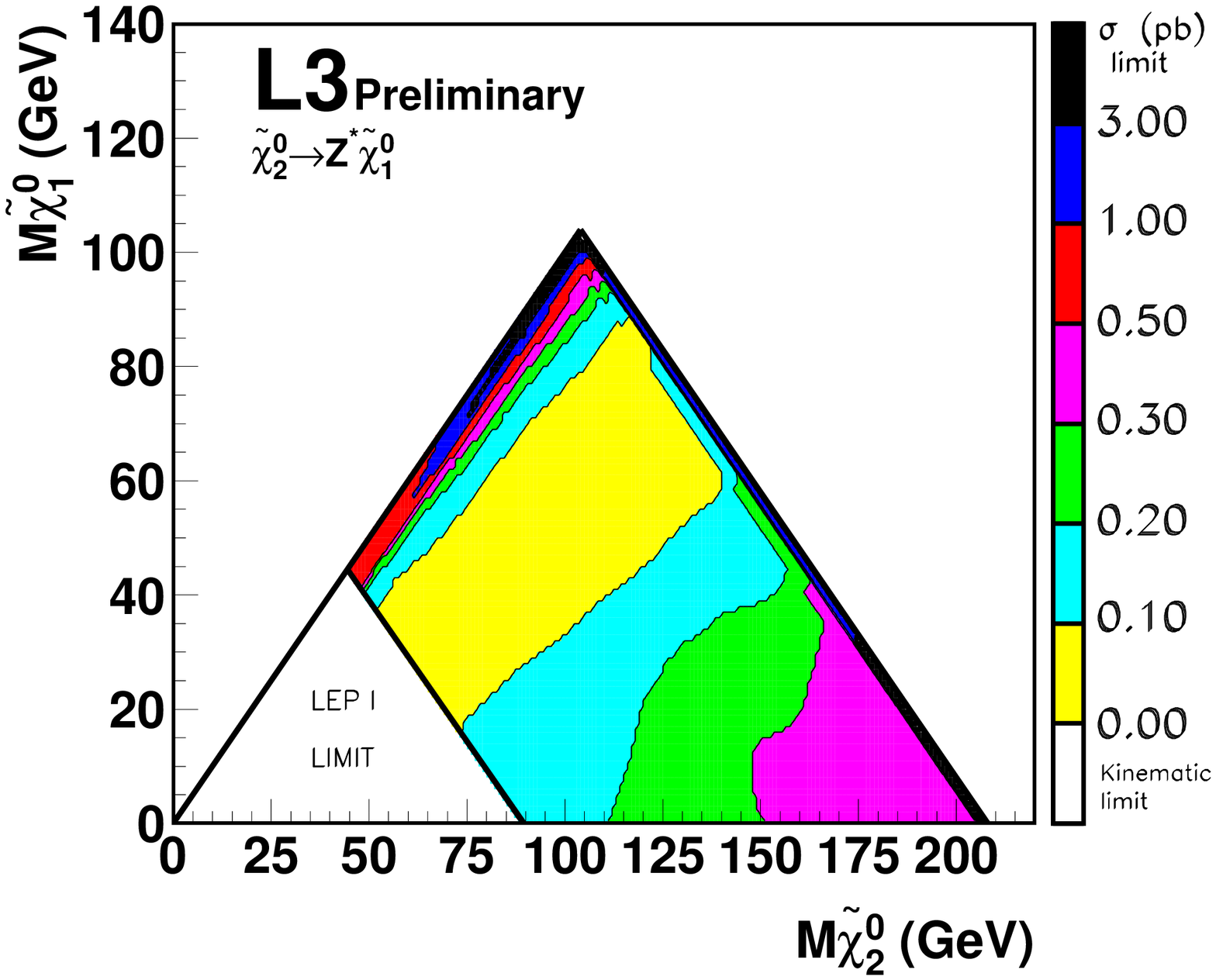}
\caption{Upper limits on neutralino production cross-sections up to 
\sqs\ = 208 \GeV . Exclusion limits are obtained assuming standard Z branching ratios
\label{fig:neucross}}
\end{minipage}
\end{figure}

All those signatures for chargino and neutralino decays have been searched for 
at LEP2 and limits on the chargino mass and production cross-section assuming 
a heavy sneutrino are shown in figures~\ref{fig:chamass} and ~\ref{fig:chacross} where \DM\ was required to be greater than 3 \GeVcc .
Figure~\ref{fig:chadeg} shows the excluded chargino mass region for low values of \DM\ (i.e.  $\DM\ <$ 3 \GeVcc) and limit on
the \XN{1}\XN{2}\ production cross-section in the plane (\MXN{1},\MXN{2}) is given in figure~\ref{fig:neucross}.

\subsection{Lightest Supersymmetric Particle mass limit}\label{subsec:char}
\begin{figure}[h]
\begin{minipage}{.49\linewidth}
The steep solid (dashed) curve  shows the effect of the
searches for the Higgs boson
for the maximal $M_{\hn}$ scenario (no mixing scenario),  
$m_0 \le $ 1000~\GeVcc\ and
$M_t$= 174.3~\GeVcc, which amounts to excluding the region of 
$\tanb<2.36 (9.7)$. The obtained limits are summarized in the following table : 
\small
\begin{center}
\begin{tabular}{|l|c|}
\hline
{Validity conditions}     & \MXN{1} limit \\
\hline
\it{high $m_0$, $\tanb >$1,incl. Higgs }&   $>$ 49 \GeVcc  \\
\hline
\it{any $m_0$, $\tanb <$40, no mixing } &  $>$ 46 \GeVcc  \\
\hline
\it{any $m_0$, $\tanb <$40, mixing, $A_{i}$=0} &  $>$ 36.7 \GeVcc  \\
\hline
\end{tabular}
\end{center}
\end{minipage}\hfill
\begin{minipage}{.49\linewidth}
\epsfysize=7.0cm\epsffile{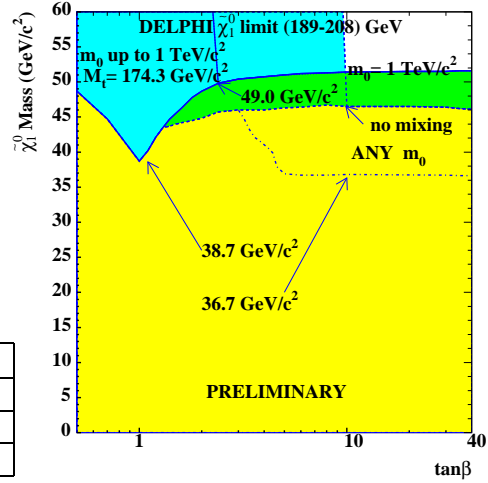}
\caption[MSSM limits in ($\mu$,$M_2$) plane]{
The lower limit at 95~\% confidence level on the mass of the lightest
neutralino, \XN{1}, as a function of \tanb\ assuming a stable \XN{1}.
 }
\label{fig:LSPLIM}
\end{minipage}
\end{figure}

The negative search results in all the channels mentioned above were used to 
exclude regions in the parameter space of the constrained MSSM and set limits on the 
LSP mass. We present here the results from the DELPHI collaboration (other LEP experiments results can be found in~\cite{results}).
Figure~\ref{fig:LSPLIM} shows the obtained \MXN{1} limit as a function of \tanb\ for different validity conditions.
The solid curve shows the limit obtained for 
$m_0$~=1000~\GeVcc, the dashed curve shows the limit obtained
allowing for  any $m_0$ assuming that there is no mixing in the
third family
($A_{\tau}=\mu\tanb$, $A_{b}=\mu\tanb$,$A_{t}=\mu/\tanb$),
and the dash-dotted curve shows the limit obtained for
any $m_0$ allowing for the mixing with 
$A_{\tau}$=$A_{b}$=$A_{t}$=0.



\end{document}